\documentclass[twocolumn,showpacs,amssymb,aps]{revtex4}

\usepackage{graphicx}
\usepackage{psfig}
\usepackage{dcolumn}
\usepackage{bm}

\def\lessim{\lower.5ex\hbox{$\; \buildrel < \over \sim \;$}}
\newcommand{\pathnow}{}
\begin{document} 
\topmargin -0.8cm
\preprint{}

\title{Strangeness Excitation Function  in Heavy Ion Collisions}

\author{Johann Rafelski}
\affiliation{Department of Physics, University of Arizona, Tucson, Arizona, 85721, USA\\
and\\
CERN-TH, 1211-Geneva 23, Switzerland}
\author{Jean Letessier}
\affiliation{Laboratoire de Physique Th\'eorique et Hautes Energies\\
Universit\'e Paris 7, 2 place Jussieu, F--75251 Cedex 05.
}

\date{August 14, 2003}

\begin{abstract}
We study as function of energy  strangeness
created in relativistic heavy ion collisions.
We consider statistical hadronization with 
chemical freeze-out in   both  equilibrium and nonequilibrium. 
We obtain strangeness per baryon and per  entropy in the  energy range  
$8.75\lessim\sqrt{s_{NN}}\lessim 200\,A$~GeV. 
A baryon density independent evaluation 
of the kaon to pion ratio is presented.
\end{abstract}

\maketitle
Production of strange hadrons  in heavy ion collisions 
has been  predicted to be  sensitive to  deconfinement. Firstly,
there is the establishment by way of gluon fusion reaction 
$gg\to s\bar s$  of an abundant  supply of strange quarks and 
antiquarks. Once the  quark--gluon plasma (QGP) cools  
to the point of hadronization,
in a second and independent step the final state
hadrons are  made in a recombination--fragmentation process~\cite{Koc86ud}.
 A specific deconfinement feature is the  enhanced  production of strange
antibaryons, increasing with strangeness content, a feature 
seen in recent experiments~\cite{Man03}.

We analyze here experimental results in search for discontinuities 
in excitation of strangeness  as function of reaction energy.
The  production yields of  (strange) hadrons   are studied 
in several experiments  at the 
 RHIC
and at the
 SPS.
Understanding of hadron yields in terms of  phase space densities 
allows us to evaluate the global properties  
of all particles produced~\cite{Raf03ma}. The particle production can be 
well described in a very large range of yields solely by evaluating
the accessible phase space size,  when including many
 hadron  resonances~\cite{Hag65}.

In this work, we consider  chemical equilibrium and non-equilibrium,
i.e., we allow  quark pair phase space 
occupancies, for  light quarks $\gamma_q\ne 1$, and/or 
 strange quarks $\gamma_s\ne 1$~\cite{Let00b} and we  require balance 
of strange and antistrange quark content~\cite{Let93hi}. 
There are two independent fit parameters when we assume complete chemical
equilibrium, the 
chemical freeze-out temperature $T$ and $\mu_b$ the baryochemical potential
(or equivalently, the quark fugacity $\lambda_q=e^{\mu_b/3T}$). 
Adding the possibility that the number of strange quark pairs is not
in chemical equilibrium, $\gamma_s\ne 1$, we  have  3 parameters, and
allowing also that light quark pair number is not in chemical 
equilibrium, we have 4 parameters. These three alternatives will be coded as open
triangles (green online), open squares (violet) and filled squares
(red), respectively,  in the results we present graphically.

Statistical hadronization cannot be 
modeled completely today, as we neither know all hadron
resonances, nor do we know the required  branching ratios of resonance
decays. This introduces arbitrariness  in
the model which  can lead to   discrepancies between
hadronization analysis results. 
To estimate the systematic error, we have performed
a study varying the pion  yield artificially  by a factor $0.8<f_\pi<1.2$.
 One can 
infer from this study that, if a relatively  large hadronization temperature
is reported, the presumption must be made that the statistical hadronization 
program used does not produce as many pions as required by  an
extrapolation of resonance mass spectrum and resonance decay pattern.  
Our computed  yields have been cross-checked (for the  chemical equilibrium 
variant) as noted in  acknowledgments.

At RHIC, the baryon yield is found to contain many
strange baryons. Not all results have been corrected for ensuing 
weak decays.  We assume in our approach that 50\% of weak decays from $\Xi$ 
to $\Lambda$ and from $\Omega$ to $\Xi$ are inadvertently included in the
yields, when these had not been corrected for such decays. We further assume that  
pions from such weak decays are {\it not} included in the experimental yields,
as these pions can clearly be shown to originate outside the interaction vertex.

We  have carried out a RHIC-200 (i.e., $\sqrt{s_{NN}}=200\,A$ GeV) Au--Au reaction
analysis based on reported  BRAHMS~\cite{Bea03fw,Ou02gm}, 
PHENIX~\cite{Adl03cb}, PHOBOS~\cite{Bac02ks},  STAR~\cite{Yam03ip,Ma03jt,Zha03dp},  
results for 
$\pi,\ h^-,\ p,\ \bar p,\ K,\ \overline{K},\ K^*,\ \phi,\ \Omega,\overline\Omega$
yields. We  have further 
reanalyzed the extensive  RHIC-130   Au--Au  results 
(compare \cite{Raf0201}) both
in order to account for latest resonance yields and  to make sure
that there is no significant change introduced by the refinements
made in the hadronization program.
For the  SPS,  we take results obtained with  Pb beams reacting with  Pb 
stationary target at  $\sqrt{s_{NN}}=8.75, 12.25,17.2$~GeV  
(projectile energy 40, 80, and 158\,$A$\,GeV). 
We use here the SPS NA49-experiment $4\pi$ particle 
multiplicity  results~\cite{Gaz03a,Gaz03b}, which include 
$\pi^\pm,\ K,\ \overline{K},\ \Lambda,\ \overline{\Lambda},\ \phi$ at
40, 80, 158$A$ GeV. We also  fit (relative) yields  of
 $\Xi,\ \overline\Xi,\ \Omega,\ \overline\Omega$ 
when available. Since we fit  $4\pi$-particle yields, no information about
the collective flow velocity is obtained.

\begin{table}[bht]
\caption{
\label{TmuRHIC} The chemical freeze-out  statistical parameters found for 
 nonequilibrium (left) and semi equilibrium (right) 
fits to RHIC results. We show $\sqrt{s_{NN}}$,  
the  temperature $T$,  baryochemical potential $\mu_b$, 
strangeness chemical potential $\mu_S$, 
the quark occupancy parameters $\gamma_q$ and
$\gamma_s/\gamma_q$, and 
the statistical significance of the fit. 
The star  (*) indicates that there is an upper limit on the value of 
$\gamma_q^2<e^{m_\pi/T}$ (on left), and/or that the value is set (on right).
}\vspace*{0.2cm}
\begin{tabular}{|l|cc|cc|}
\hline
$\sqrt{s_{NN}}$\,[GeV]    &200          & 130        &200          & 130    \\[0.1cm]
\hline
$T$\,[MeV]                &$143\pm7$    & $144\pm3$  &$160\pm8$    & $160\pm4$\\[0.1cm]
\hline
$\mu_b$\,[MeV]            &$21.5\pm1$&$29.2\pm1.5$   &$24.5\pm1$   &$31.4\pm1.5$ \\[0.1cm]
$\mu_S$\,[MeV]            &$4.7\pm0.4$ &$ 6.6\pm0.4$ &$5.3\pm0.4$ &$ 6.9\pm0.4$ \\[0.1cm]
\hline
$\gamma_q$                &$1.6\pm0.3^*$&$1.6\pm0.2^*$& $1^*$ & $1^*$ \\[0.1cm]
$\gamma_s/\gamma_q$       &$1.2\pm0.15$ & $1.3\pm0.1$ &$1.0\pm0.1$&$1.13\pm0.06$ \\[0.1cm]
\hline
\hline
$\chi^2/$dof               &2.9/6        & 15.8/24    &4.5/7       & 32.2/25  \\[0.1cm]
\hline
\end{tabular}
 \end{table}

 We present the RHIC freeze-out statistical 
parameters in  table \ref{TmuRHIC}. These results are
incorporating statistical and systematic errors
in the data, along with errors in the statistical 
hadronization theory,  arising from the above described 
uncertainty  of the pion yield. 
The bottom line in  table \ref{TmuRHIC} 
presents the statistical significance. 
The introduction of the full  chemical nonequilibrium (on left)
reduces by factor two the  value of $\chi^2$. 
Even when allowing for all
systematic uncertainties and theoretical uncertainty regarding
the  pion yield, the nonequilibrium fit is a more compelling considering
 the (presently) more data rich
RHIC-130 system.  The confidence level of the chemical 
equilibrium approach remains here  at 15\%. 
At RHIC-200, the current data can be interpreted within
a chemical equilibrium model,  since the decisive
results on relative yields of (multi)strange 
(anti)hyperons  are not yet available.

The  statistical hadronization is expected to describe  particle
production well in presence of  a  sudden QGP
breakup. Statistical hadronization approach 
is not appropriate when there is a lot of
hadron--hadron rescattering in the final state, which
may be the case at low SPS energies and below. 
In  this  case,  kinetic models need to 
be applied, introducing  a multitude of freeze-out conditions
depending on the nature of particle considered. Accordingly, 
 when we fit the low energy SPS results, a less favorable $\chi^2$ 
is found than at RHIC, or at the top SPS energy. However, 
the stability of the physical properties we extract when we subject the 
system to perturbations regarding pion yield indicate that results
concerning the physical properties of the hadron source 
are reliable.

\begin{figure}[t]
\vskip 0.3cm
\hspace*{-.2cm}\psfig{width=8.5cm,figure=\pathnow  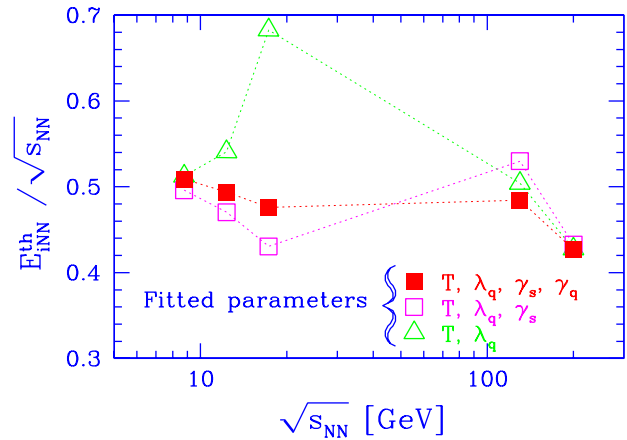}
\caption{\label{PLEBS}
Fraction of energy stopping at SPS and RHIC: 
results are shown for 40, 80, 158$A$ GeV Pb--Pb,
  200$A$ GeV S--W/Pb reactions and at RHIC for  65+65$A$ GeV Au--Au
interactions. Connecting lines guide the eye (color online). 
}
\end{figure}

With the statistical parameters fixed, the properties of the fireball
can be computed, evaluating the contributions made by  each of the hadronic 
particle species. We  first show how this works considering the  energy stopping
 in figure \ref{PLEBS}. This result allows us to  
 represent below  our further findings  as function of 
 $E_{iNN}^{\rm th}$, the specific
per baryon pair thermal energy available at the time of hadronization. This
dependence substitutes for the dependence on $\sqrt{s_{NN}}$,
the initial energy per baryon pair 
brought into the reaction. We see in figure \ref{PLEBS}, that  
the chemical equilibrium fit (open triangles) shows counterintuitive 
behavior.

We note that  the 
fraction of the  energy per baryon pair  which is initially thermalized  
is obtained from this result by adding the kinetic energy of the
 collective matter flow.  Using at RHIC-200 as the 
average transverse flow velocity $\langle v_\bot^2\rangle =0.45^2$ 
(and smaller values at smaller collision energies) we 
find a $\simeq 20$\% correction. Thus, we see that  $50\pm5$\%
of the energy per baryon carried into the reaction is initially made available to
the thermal degrees of freedom, and this value is 
independent of the  collision energy in the entire SPS and  RHIC range
$9<\sqrt{s_{NN}}<200$ GeV. This means that the baryon stopping is two times
larger then energy stopping.

We are now ready to consider the final state 
specific per baryon  strangeness  yield. 
We  evaluate strangeness  yield $s$, the number of produced  strange quark pairs 
and divide   it by the similarly computed 
thermal fireball  baryon number content $b$. $b$  is obtained 
summing the net (particle minus antiparticle) yields
of all nucleons, (multi strange) hyperons 
and their resonances. The baryon number  $b$ is 
naturally conserved. Strangeness is predominantly  produced in the 
early stage of the reaction, when the density and temperature are
highest, and there is little change in this yield during the late fireball evolution. 
Thus, $s/b$  ratio probes directly  the extreme initial conditions. 
Consideration of this yield ratio eliminates the
absolute  yield normalization parameter (sometimes but not always 
`reaction' volume), as well as some  uncertainties originating 
in the experimental particle yield data, which propagate through 
the analysis of experimental data.

\begin{figure}[!thb]
\hspace*{-.40cm}\psfig{width=8.5cm,figure=\pathnow  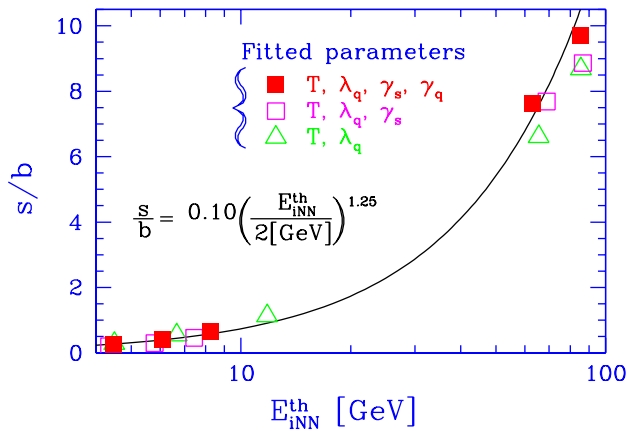}
\caption{\label{PLETRB}
Strangeness per thermal  baryon  participant, $s/b$  as function of 
thermal specific energy content $E^{\rm th}_{{\rm i}\,NN}$. 
}
\end{figure}

In figure \ref{PLETRB}, we show $s/b$ as function of the thermal specific energy 
content $E^{\rm th}_{{\rm i}\,NN}$. Strangeness  yield is continuous in the
entire energy domain as is shown by the solid line drawn to guide the eye. 
The rise of specific strangeness yield is slightly faster than linear with energy, 
as is indicated in the  figure insert. (The appearance of an exponential shape
is due to logarithmic energy scale in  figure~\ref{PLETRB}). 
The results for SPS energy range are in quantitative agreement with 
predictions made  assuming a QGP state of matter and gluon fusion
strangeness formation mechanism, compare  figure 38 in Ref.~\cite{Acta96}. 
The SPS specific strangeness  yield extrapolates smoothly  to the RHIC energy 
range. The reaction mechanism 
producing strangeness and stopping baryon number are evolving in parallel 
yielding a smooth change in the ratio of both variables.

The highest energy $\sqrt{s_{NN}}$=200 GeV data point 
in figure \ref{PLETRB} seems to be  slightly
lower than the presented extrapolation predicts. We think that this yield  will
increase once we have included in the statistical hadronization analysis the hyperon 
yield ratios $\Xi/\Lambda, \Lambda/p$. The presence, in the global fit, of these particle
ratios will increase the fitted strangeness yield, provided that their measured 
values  are  comparable in magnitude to those reported at  $\sqrt{s_{NN}}$=130 GeV.

An important feature of the RHIC experimental results, shown
 in figure \ref{PLETRB},
is the large  strangeness yield per participating baryon.
Although it has been early on noted that at RHIC-130, up to 8 strange 
quark pairs per baryon are produced~\cite{Raf01hp}, little attention has been
given to this high yield. At RHIC-2000 each interacting 
baryon pair  produces about 20  strange quark--antiquark pairs.

\begin{figure}[!thb]
\hspace*{-.40cm}\psfig{width=8.5cm,figure=\pathnow  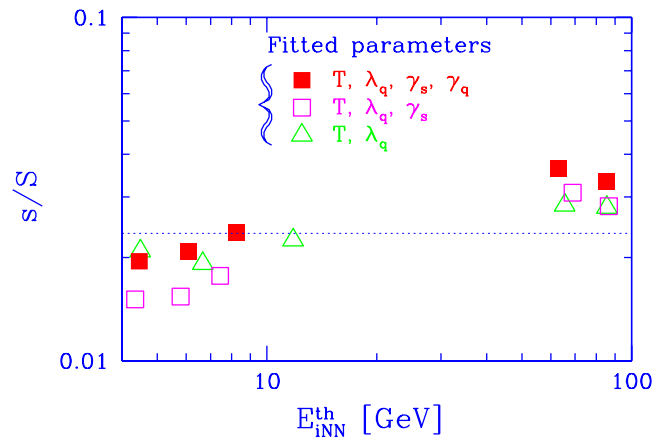}
\caption{\label{PLETRBSLOG}
Strangeness per  
entropy, $s/S$, as function of the thermal specific energy 
content $E^{\rm th}_{{\rm i}\,NN}$.
}
\end{figure}

An interesting question is how this great increase in yield
compares to the production of particles in general. The 
Wr\'oblewski ratio
$
W_s\equiv 2 \langle s\bar s\rangle/\langle u\bar u +d \bar d\rangle
$
 is often used in such a comparison. Recently it has
been  realized that this ratio can be artificially enhanced by high baryon 
density which can suppress light quark pair yields~\cite{Br01as}. 
Therefore we here compare
the strangeness production to the global entropy $S$ yield. 
We evaluate $S$ in the same
way as we have obtained other global properties. 
There is an additional nonequilibrium  entropy term  to
be allowed for when chemical non-equilibrium prevails. 
Both entropy $S$ and strangeness $s$ 
are nearly conserved in the  hydrodynamical expansion of QGP and 
increase only moderately in the hadronization of QGP. 
The observed ratio $s/S$ is  established by  microscopic reactions
operational in the early stages of the heavy ion 
collision.  

Strangeness per entropy, $s/S$, is presented in
 figure \ref{PLETRBSLOG} as function of the specific thermal energy content.
The horizontal line is the maximum SPS yield base. 
We note  the  modest smooth rise  in  
the SPS energy domain. Most interestingly,   at RHIC, as   compared to SPS,  
an unexpected 50\% increase in $s/S$  is noted,
allowing for chemical nonequilibrium. The excitation of
strangeness  seems to rise faster with energy than the production of 
entropy. A possible  explanation of this phenomenon is that the 
hot initial state, in which the 
threshold in energy for strangeness formation has been overcome, 
lives longer when formed at RHIC conditions. However, 
it appears important to fill the energy range between SPS and RHIC 
with data in search of a new physics energy threshold.

\begin{figure}[!thb]
\hspace*{-.40cm}\psfig{width=8.7cm,figure=\pathnow  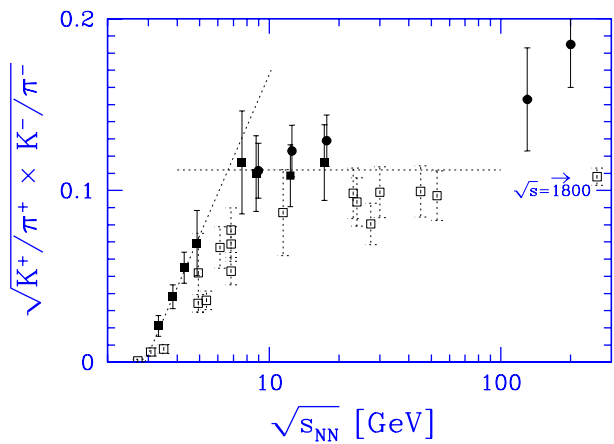}
\vskip -0.2cm
\caption{\label{Kpisqrt}
$K/\pi$ ratio as function of collision energy. 
Filled symbols are for nuclear and open for  $K^+/\pi^+$ 
measured in elementary $pp$ collisions. Squares denote the  full 
multiplicity ratio and circles the  central rapidity yield 
ratio.
}
\end{figure}

Based on the study of the $K^+/\pi^+$ particle yield ratio~\cite{Gaz03b}, 
such a new physics energy threshold had been expected  below 40$A$ GeV, 
at  the lowest SPS energy, i.e., below the energy range we could
consider in the global hadron freeze-out analysis.  Inspecting 
available particle yields,  we note that their dependence on the unknown 
baryon density is significant, mimicking new physics. This baryon density 
effect can be greatly reduced considering  the product of particle and
antiparticle yields. In this case, the chemical potential cofactors 
in particle yields cancel, being  inverse of each other, and the baryon
density effect largely disappears. To study kaon to pion ratio we thus form:
$
{K}/{\pi}\equiv \sqrt{  ({K^+}{K^-})/({\pi^+}{\pi^-})}
$,
and effectively `remove' baryon density effect present in the individual
ratios $K^+/\pi^+$ and $K^-/\pi^-$. 
Using the NA49 data set~\cite{Afa2002mx},  we present the
$K/\pi$  ratio  in figure \ref{Kpisqrt}. The filled squares are for
the 4-$\pi$ full phase space data set. The filled circles are for 
the central rapidity results, including RHIC-130 and  
RHIC-200 data  \cite{Ou02gm}. The 
central rapidity NA49 results are from figure 6 in \cite{Afa2002mx}. The 
$pp$ charged $K^+/\pi^+>K/\pi$ background is shown by open 
squares, and we indicate for 
$\sqrt{s}={1800}$ the $p$--$\bar p$ {\small TEVATRON} point \cite{Ale93wt}
which is at an energy beyond the range considered.

The lines,  in figure \ref{Kpisqrt}, guide the eye to the  trends of the 
alternate gradient synchrotron  (AGS) and SPS results.  These trends 
of behavior  intersect within the SPS low energy range 
below the lowest NA49 SPS point which was obtained at 30$A$GeV. However,
this appears here to be a smooth transition from a rise to saturation of the 
$K/\pi$  production. There is a clear enhancement of 
the $K/\pi$  ratio at RHIC compared both with the AGS/SPS trend and with
the  $K^+/\pi^+> K/\pi$  measured in elementary $pp$ collisions. This
enhancement coincides with the specific strangeness per entropy, 
$s/S$ enhancement seen in figure~\ref{PLETRBSLOG}.

We have shown that the strangeness per baryon excitation
 in  the fireball of dense matter
formed at SPS and RHIC energies is a smooth function of energy, but we 
find a step-up in strangeness per entropy between the SPS and RHIC 
energy ranges, also visible in the rise of the  $K/\pi$ ratio. Thus there are 
two energy domains to investigate for threshold behavior, the 35+35 GeV RHIC 
range and the 20$A$ GeV on fixed target low SPS energy, 
where the $K/\pi$ ratio saturates.
We further noted the quantitative agreement in the 
strangeness yield with predictions made
for the SPS energy range, assuming QGP production mechanisms. 
We have shown that  20 strange--antistrange quark pairs are made
per colliding baryon pair at the top  RHIC energy. 

{\it Note added}
More technical details 
are now available in Ref.~\cite{Zak03}. A systematic study of SPS results 
by a yet another group has just appeared \cite{Bec03}, and the numerical results
shown there agree with our SPS results. Fig 10 shows that $\gamma_q=1$ is actually
a local fit maximum for the  top SPS energy.  

{\it Acknowledgments:\/} 
We thank M. Ga\'zdzicki and M. van Leeuwen for NA49 data and results. 
We thank W. Florkowski and D. Magestro 
for  comparison of hadronization code results and fruitful discussions.
Work supported in part by a grant from the U.S. Department of
Energy,  DE-FG03-95ER40937\,. LPTHE, Univ.\,Paris 6 et 7 is:
Unit\'e mixte de Recherche du CNRS, UMR7589.\\


\vskip 0.3cm

\end{document}